\documentclass[
reprint,
superscriptaddress,
amsmath,
amssymb,
aps,
pra
]{revtex4-2}
\usepackage{graphicx}
\usepackage{dcolumn}
\usepackage{bm}
\usepackage[bookmarks=true, colorlinks=true, linkcolor=blue, urlcolor=blue, citecolor=blue]{hyperref}
\usepackage{graphicx}
\usepackage{xcolor}
\usepackage{tikz}
\usepackage{color}
\usepackage{amsfonts,amssymb}

\usepackage[ruled,linesnumbered]{algorithm2e}

\begin{document}
\preprint{APS/123-QED}

\title{Machine-Learning Insights on Entanglement-trainability Correlation of Parameterized Quantum Circuits}

\author{Shikun Zhang}
\affiliation{School of Physics, Beihang University, Beijing 100191, China}

\author{Yang Zhou}
\email[Contact author: ]{yangzhou9103@buaa.edu.cn}
\affiliation{School of Physics, Beihang University, Beijing 100191, China}
    
\author{Zheng Qin}
\affiliation{School of Physics, Beihang University, Beijing 100191, China}

\author{Rui Li}
\affiliation{School of Physics, Beihang University, Beijing 100191, China}

\author{Chunxiao Du}
\affiliation{School of Physics, Beihang University, Beijing 100191, China}

\author{Zhisong Xiao}
\affiliation{School of Physics, Beihang University, Beijing 100191, China}
\affiliation{School of Instrument Science and Opto-Electronics Engineering, Beijing Information Science and Technology University, Beijing 100192, China}

\author{Yongyou Zhang}
\email[Contact author: ]{yyzhang@bit.edu.cn}
\affiliation{School of Physics, Beijing Institute of Technology, Beijing 100081, China}

\begin{abstract}
Variational quantum algorithms (VQAs) have emerged as the leading strategy to obtain quantum advantages on the current noisy intermediate-scale devices. 
However, their entanglement-trainability correlation indicates that deep circuits are generally untrainable due to the barren plateau (BP) phenomenon, which challenges their applications.
In this work, we suggest a gate-to-tensor (GTT) encoding method for parameterized quantum circuits (PQCs), with which two long short-term memory networks (L-G networks) are trained to predict both entangling capability and trainability. 
The remarkable capabilities of L-G networks afford a statistical way to investigate the entanglement-trainability correlation of PQCs within a dataset encompassing millions of generated circuits.
This machine-learning-driven method first confirms the negative correlation between entanglement and trainability.
Then, we observe that circuits with any values of entangling capability and trainability exist. These circuits with high entanglement and high trainability possess an appropriate high portion of nearest-neighbor non-local gates.
Furthermore, the trained L-G networks can be employed to construct PQCs with specific entanglement and trainability, demonstrating their practical applications in VQAs.
\end{abstract}	

\maketitle

\section{Introduction}   

In the current noisy intermediate-scale quantum era \cite{1-1, 1}, variational quantum algorithms (VQAs) garner wide interest due to their low consumption of quantum resources and noisy friendliness \cite{0, 7}.
Reported VQAs include variational quantum eigensolver \cite{2-4-ent,2-5-ent,2-6-ent,2-7-ent,2-8-ent,2-9-ent,qinzheng}, quantum approximate optimization algorithm \cite{2-10-ent,2-11-ent,2-12-ent}, variational quantum machine learning \cite{2-15-ent,2-17-ent,2-19-ent} and so on.
These VQAs are implemented by parameterized quantum circuits (PQCs).
PQCs primarily determine the computational performance of VQAs \cite{8-ent, zsk}.
Given that entanglement is the foundation of the unique characteristics and advantages of quantum systems over classical ones \cite{lirui,lirui-1,lirui-2,7-ent,7-34-ent,8-ent,18}, it naturally serves as a primary guiding principle in constructing PQCs. However, the heuristic nature of VQAs results in a lack of guaranteed performance of PQCs \cite{7-ent}.
Significant researches on VQAs' performance have been reported \cite{7-ent,7-24-ent,7-25-ent,7-26-ent,7-27-ent,7-28-ent,7-29-ent,7-30-ent, 7-31-ent,7-32-ent,7-33-ent,7-34-ent,7-35-ent}.
On the one hand, PQCs with high entanglement were required to capture non-trivial quantum correlations or ground states of interested quantum systems \cite{5,5-8,5-9,5-20}. On the other hand, the improper amount of entanglement may hinder the performance of VQAs \cite{8-ent, 8-48-ent}. A notorious issue is the barren plateau (BP) phenomenon \cite{bp-original, 18, ENTvsBP-24}.

When BP occurs, the gradients of the cost functions in VQAs vanish exponentially with the system size \cite{EXPvsBP-28,EXPvsBP-29,7-27-ent, EXPvsBP-31,EXPvsBP-32}. 
This implies an exponential number of measurement shots are needed to resolve and determine the cost-minimizing direction. 
Such exponential complexity would undermine the quantum advantages of VQAs \cite{7}. 
So far, some evidence \cite{ENTvsBP-24, 18} suggests a rudimentary negative correlation between entanglement and trainability.
Nevertheless, the more fundamental and general relationship between them remains elusive.
The main challenge lies in the enormous number of possible PQCs and the practical difficulties in evaluating their entangling capability and trainability.
Fortunately, machine learning has shown great potential in addressing these complex issues. Current research has explored their applications in performing quantum architecture searchers \cite{2-ent}, constructing generative quantum eigensolvers \cite{GQE}, synthesizing quantum circuits \cite{QCdiffusionModel} and predicting PQCs' expressibility \cite{learnExp1,learnExp2}. In addition, several practical quantum machine learning algorithms have recently emerged, including quantum ghost imaging \cite{PRARef1}, quantum metrology based on reinforcement learning \cite{PRARef2}, trainable generative quantum models \cite{PRARef3}, and expressivity-enhanced quantum neural networks \cite{PRARef4}. These developments underscore the growing synergy between machine learning and quantum computing, offering promising strategies to address persistent challenges in quantum algorithms.

 \begin{figure*}
\centering
\includegraphics[width=0.99\textwidth]{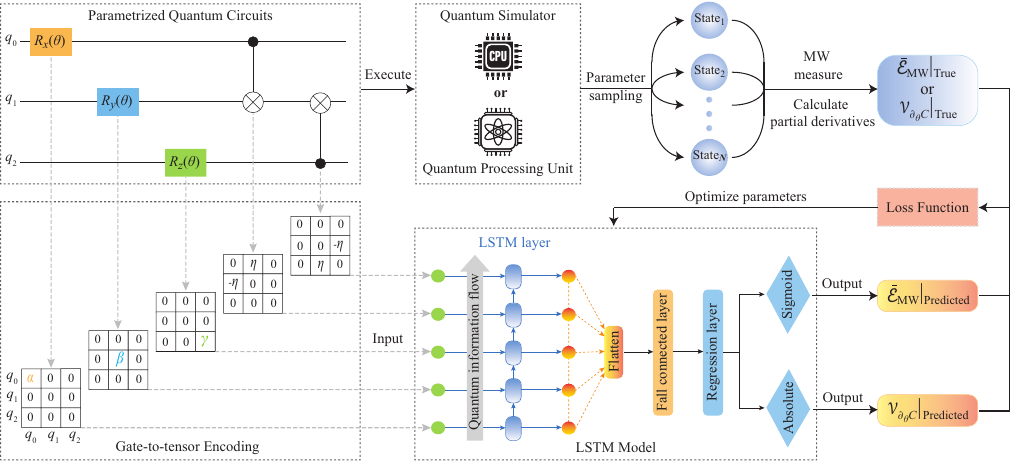}
  \caption{The L-G networks for predicting the entangling capability (${\bar {\cal E}}_{\rm MW}$) and trainability (${\cal V}_{\partial_\theta C}$) of PQCs. The true ${\bar {\cal E}}_{\rm MW}$ or ${\cal V}_{\partial_\theta C}$ of PQCs is determined by sampling the parameter space multiple times and executing PQCs on a quantum simulator or quantum processing unit. We utilize the proposed gate-to-tensor (GTT) method to encode PQCs into a tensor. This tensor is then fed into the LSTM model across different time steps to predict the ${\bar {\cal E}}_{\rm MW}$ and ${\cal V}_{\partial_\theta C}$.}
 \label{fig1}
\end{figure*}

In this work, we propose a machine-learning method to efficiently evaluate entangling capability and trainability based on which the general entanglement-trainability correlation of PQCs is explored.
An innovative scheme is suggested to encode PQCs into tensors, referred to as gate-to-tensor (GTT) encoding.
Using the GTT code as input, two long short-term memory (LSTM) networks, dubbed the L-G network, are trained to predict both entanglement and trainability, as illustrated in Fig.~\ref{fig1}.
The L-G network provides three key advantages:
(i) Its one-gate-to-one-tensor approach ensures consistent tensor dimensions for PQCs with equal numbers of quantum gates and qubits;
(ii) The GTT encoding effectively captures the mutual control information among qubits;
(iii) The LSTM block's temporal architecture naturally models information flow in PQCs.
The remarkable capabilities of the L-G networks afford a statistical way to delve into the entanglement-trainability correlation within a dataset encompassing millions of generated circuits.
Our data-driven method first confirms the negative correlation between trainability and entanglement. It is further demonstrated that this negative correlation is `statistical' rather than determinative. There still exist circuit structures with both high entanglement and high trainability. Based on these findings, we establish three key topological parameters for PQCs: C-Not gate ratio, connectivity density, and average adjacency matrix, which collectively characterize these optimal structures.
Finally, by utilizing the discovered statistical properties of the three topological parameters and the trained L-G networks, we develop an algorithm for constructing PQCs with tailored entanglement and trainability properties.
 This algorithm achieves a million-fold improvement in time efficiency, demonstrating significant potential for practical applications in variational quantum algorithms.

\section{Preliminary}

\subsection{Parameterized Quantum Circuits}

PQCs act as a bridge connecting classical and quantum computing \cite{5}. Their output states can be written as
\begin{equation}
  \vert\psi_{\boldsymbol\theta}\rangle = U(\boldsymbol\theta)\vert0\rangle^{\otimes{n}},
  \label{equ3}
\end{equation}
where $U(\boldsymbol\theta)$ is a parameterized unitary operator, transforming the $N$-qubit reference state $\vert0\rangle^{\otimes n}$ into the output one $\vert\psi_{\boldsymbol\theta}\rangle$. 
Adjusting the parameters $\boldsymbol\theta$ leads to different output states. 

The choice of PQCs depends on the target tasks and therefore, various ansatz architectures for PQCs are developed \cite{5-2, 5-17, 5-18, 5-19}. 
In this work, we adopt the hardware-efficient ansatz (HEA) since it is constructed by gates that can be directly executed on near-term quantum hardware. 
A typical HEA consists of single-qubit gate layers with tunable parameters and two-qubit gate layers providing entanglement. 
Its parameterized unitary operator, accordingly, has the form,
\begin{eqnarray}
  U(\boldsymbol\theta) =\prod_{l=1}^{\mathbb{L}
}U_{l}(\theta_{l})W_{l},
  \label{equ2}
\end{eqnarray}
with 
\begin{eqnarray}
  U_{l}(\theta_{l}) = \bigotimes_{j=1}^{N}R_{\alpha}(\theta_{l}^{j}),
  \label{equ3}
\end{eqnarray}
where $R_{\alpha}(\theta_{l}^{j}) = e^{-i\theta_{l}^{j}\sigma_\alpha/2}$ and $\sigma_\alpha$ is the Pauli matrix with $\alpha = x,\ y,\ {\rm or}\ z$. 
The total number of qubits is denoted by $N$. 
$W_{l}$ represents the operator of the $l$th unparametrized two-qubit gate, with total number being $\mathbb{L}$.

\begin{figure}
\centering
\includegraphics[width=0.49\textwidth]{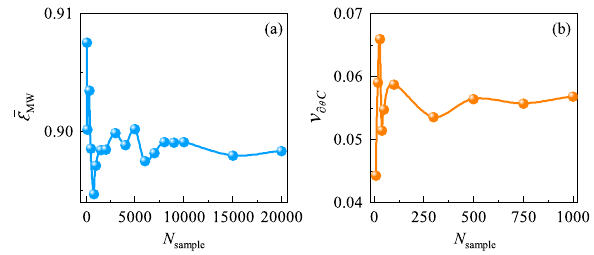}
  \caption{Statistical variations of the entangling capability ${\bar {\cal E}}_{\rm MW}$ (a) and trainability ${\cal V}_{\partial_\theta C}$ (b) of the 8-qubit PQCs with the number of samples $N_{\rm sample}$. As $N_{\rm sample}$ increases, the ${\bar {\cal E}}_{\rm MW}$ and ${\cal V}_{\partial_\theta C}$ tend to converge and the confidence level increases.}
  \label{SI-fig1}
\end{figure}

\subsection{Entangling capabilities}\label{Ent}

The entangling capability of a PQC is linked to its average ability of creating entanglements and is commonly measured by state entanglement. Here, we utilize the sampling average of the Meyer-Wallach (MW) entanglement measure \cite{5-33} to quantify the entangling capability of a PQC. For the $N$-qubit system, the MW entanglement, ${\cal E}_{\rm MW}$, is defined as below:
\begin{equation}
  {\cal E}_{\rm MW}(|\psi\rangle) \equiv \frac{4}{n}\underset{i=1}{\overset{n}{\sum}} \mathcal{D}(\Gamma_{i}(0)|\psi\rangle, \Gamma_{i}(1)|\psi\rangle),
\end{equation}
where $\Gamma_{i}(b)$ is a linear mapping that acts on a computational basis with $b\in\lbrace0, 1\rbrace$. 
That is,
\begin{equation}
\Gamma_{i}(b)|b_{1}\dots{b_{n}}\rangle = \delta_{bb_{i}}|b_{1}\dots{\overset{\circleddash}{b}_{i}}\dots{b_{n}}\rangle,
\end{equation}
where the symbol $\circleddash$ means to remove the $i$-th qubit. The $\mathcal{D}$ operation is the generalized distance, defined as: 
\begin{equation}
\mathcal{D}(|u\rangle, |v\rangle) = \frac{1}{2}\underset{i,j}{\sum}|u_{i}v_{j}-u_{j}v_{i}|^2   
\end{equation}
with $|u\rangle = \sum{u_{i}|i\rangle}$ and  $|v\rangle = \sum{v_{i}|i\rangle}$.
 
As a global measure of multi-particle entanglement for pure states, the MW measure has been widely employed as an effective tool in various quantum information applications \cite{5-33,5-37}, offering insights into entanglement properties. 
It is particularly suitable for quantifying the entangling capabilities of PQCs by evaluating their output states \cite{5}.
In detail, the entangling capability of a PQC is measured by the average MW entanglement (${\bar {\cal E}}_{\rm MW}$) among the PQC's output states, sampled in the parameter space of $\bm \theta$, i.e.,
\begin{equation}
  {\bar {\cal E}}_{\mathrm{MW}} \equiv \frac{1}{|S|}\underset{\boldsymbol\theta_{i}\in{S}}{\sum}\mathrm{MW}(|\psi_{\boldsymbol\theta_i}\rangle),
  \label{equ7}
\end{equation}
where $S = \lbrace\theta_i\rbrace$ represents the sets of sampled parameter vectors in the parameter space of a PQC and $|S|$ represents the number of sampled parameter vectors. We must sample the parameter space sufficiently to ensure the convergence of ${\bar {\cal E}}_{\rm MW}$ calculated by Eq.(\ref{equ7}). Fig.~\ref{SI-fig1}(a) indicates that $|S|$ should be on the order of twenty thousand.

\subsection{Trainability}

Trainability is an important factor that influences the performance of PQCs. 
Optimization issues such as the BP have attracted plenty of attention \cite{16, 16-13, 16-14, 16-15, 16-16}. 
The so-called BP phenomenon refers to the gradient of a cost function that vanishes exponentially with the system size.
We follow the Ref.~\cite{bp-original} to measure the trainability of a PQC.
Firstly, $U(\boldsymbol\theta)$ in Eq.~\eqref{equ2} is separated into the left and right parts,
\begin{equation}
U(\boldsymbol\theta) = U_{\mathcal{L}}(\boldsymbol\theta)U_{\mathcal{R}}(\boldsymbol\theta),
  \label{equ4}
\end{equation}
where 
\begin{eqnarray}
U_{\mathcal{L}}(\boldsymbol\theta) =\prod_{l=k+1}^{L}U_{l}(\theta_{l})W_{l},
\ \ \ U_{\mathcal{R}}(\boldsymbol\theta) = \prod_{l=1}^{k}U_{l}(\theta_{l})W_{l}.
\end{eqnarray}
Secondly, the cost function ($C$) is defined as
\begin{equation}
C = Tr[HU(\boldsymbol\theta)\rho{U(\boldsymbol\theta)^{\dagger}}],
\end{equation}
where $H$ is a Hermitian operator and $\rho$ is an initial state.
Thirdly, the gradient of $C$ with respect to $\theta_{k}$ can be written as \cite{bp-original}:
\begin{equation}
\partial_{\theta_{k}}C\equiv\frac{\partial{C(\boldsymbol\theta)}}{\partial\theta_{k}}=i\langle0\vert{U^{\dagger}_{\mathcal{R}}}\lbrack{V_{k}},U^{\dagger}_{\mathcal{L}}HU_{\mathcal{L}}\rbrack{U_{\mathcal{R}}}\vert{0}\rangle,
\label{equ11}
\end{equation}
with the Hermitian operator $V_{k}=\bigotimes_{j=1}^{n}\sigma_{j}$.
Finally, the variance of $\partial_{\theta_{k}}C$ is used to measure the trainability, defined as
\begin{equation}
{\cal V}_{\partial_\theta C} = \langle(\partial_{\theta_{k}}C)^{2}\rangle.
\label{equ12}
\end{equation}
It should be pointed out that the ${\cal V}_{\partial_\theta C}$ is strongly influenced by the choice of the cost function \cite{EXPvsBP-28}. While global cost functions typically suffer from BP with exponentially vanishing gradients in system size \(n\), local cost functions demonstrate more favorable scaling. Their gradients decay polynomially with \(n\) when circuit depth scales logarithmically \cite{EXPvsBP-28,yuanxiao-198,yuanxiao-199}. For robust analysis of the ${\bar {\cal E}}_{\rm MW}-{\cal V}_{\partial_\theta C}$ relationship, we adopt a 2-local cost function with \(H = \sigma_{1}^{z}\sigma_{2}^{z}\). This choice capitalizes on local circuit properties while circumventing the gradient suppression inherent to global cost functions. We must sample the parameter space sufficiently to ensure the convergence of ${\cal V}_{\partial_\theta C}$ calculated by Eq.(\ref{equ12}). Fig.~\ref{SI-fig1}(b) shows that the sampling number is on the order of one thousand.

\section{Method}

\subsection{Dataset}

For training the L-G networks, we generate all 8-qubit PQCs through random configurations of gate types and qubit assignments, with gate counts varying uniformly between 30 and 50. 
This setting ensures PQCs with rich topological connectivity and CNOT gate ratios approximately spanning from 0 to 1, allowing for broad variations in ${\bar {\cal E}}_{\rm MW}$ and ${\cal V}_{\partial_\theta C}$. It enables a comprehensive exploration of their relationship to achieve the research objectives.
To accommodate the variability in circuit depth, we implement GTT encoding with $L=50$, padding shorter circuits with zero-filled $N\times N$ matrices.  We generated fifty thousand PQCs and divided them into training and testing datasets in a ratio of $9:1$.

The full dataset construction also incorporates the labeling procedure, which requires the calculation of ${\bar {\cal E}}_{\rm MW}$ and ${\cal V}_{\partial_\theta C}$ for each generated PQC. The computational framework is schematically illustrated in the first row of Fig.~\ref{fig1}.
The \({\bar {\cal E}}_{\rm MW}\) label data is calculated using Eq.~(\ref{equ7}) while the \({\cal V}_{\partial_\theta C}\) label data is generated via Eq.~(\ref{equ12}). 
However, achieving convergence for these quantities requires extensive sampling: approximately twenty thousand parameter configurations for ${\bar {\cal E}}_{\rm MW}$ and one thousand for ${\cal V}_{\partial_\theta C}$ as demonstrated in Fig.~\ref{SI-fig1}.
It is this observation that motivates our development of a neural network architecture capable of learning the mapping between PQC structures and their corresponding ${\bar {\cal E}}_{\rm MW}$ and ${\cal V}_{\partial_\theta C}$ values.

\subsection{L-G networks}

The L-G network architecture for predicting ${\bar {\cal E}}_{\rm MW}$ and ${\cal V}_{\partial_\theta C}$ is shown in the second row of Fig.~\ref{fig1}.
It consists of five key components: GTT encoding for input processing, LSTM layers for feature extraction, a flattening layer for feature integration, fully connected layers for dimensionality reduction, and a regression layer for final prediction. 
Detailed architectural settings are provided in Appendix~\ref{LGStrcture}.

Tensor-based encoding of PQCs serves as the foundational technology for machine learning applications in quantum circuit analysis. While existing image encoding (IE) strategies \cite{2-ent} transform PQCs into multi-channel images, they fail to capture crucial mutual control information among qubits.
To address this limitation, we introduce the GTT encoding scheme, which represents PQCs as $L\times N\times N$ tensors, where $L$ and $N$ denote the numbers of quantum gates and qubits, respectively.
This encoding scheme can accurately capture all structural information of PQCs and simultaneously be compatible with the data processing characteristics of machine learning models.
As demonstrated in Appendix~\ref{gttvsie}, the GTT encoding consistently outperforms conventional IE approaches, particularly in predicting ${\bar {\cal E}}_{\rm MW}$ through the L-G network.
Our implementation considers four fundamental quantum gates: $R_x(\theta)$, $R_y(\theta)$, $R_z(\theta)$, and C-Not.
The tensor representation distinguishes these gates through specific non-zero elements. Rotating gates have one non-zero element in their tensors, while the C-Not gate has two.
If $R_x(\theta)$ [$R_y(\theta)$ or $R_z(\theta)$] rotates the $n$th qubit, the non-zero element is the $n$th diagonal one, set to $\alpha$ [$\beta$ or $\gamma$].
If the C-Not gate represents the control of the $n$th qubit on the $n'$th, the two non-zero elements are the $nn'$th and $n'n$th, set to $\eta$ and $-\eta$, respectively.
An example of the GTT encoding with $L=5$ and $N=3$ is shown on the left of Fig.~\ref{fig1}.

\subsection{Performance}

\begin{figure}
\centering
\includegraphics[width=0.49\textwidth]{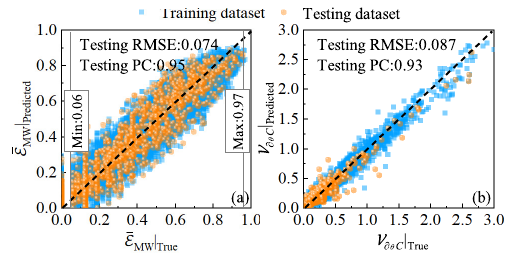}
  \caption{Confusion scatters for predicted and true values of ${\bar {\cal E}}_{\rm MW}$ (a) and ${\cal V}_{\partial_\theta C}$ (b) on both training and testing datasets with 8 qubits and 30-50 quantum gates.}
  \label{fig2}
\end{figure}

The performance of the GTT encoding depends on the encoding values of $\alpha$, $\beta$, $\gamma$, and $\eta$.
In the training process, these parameters are treated as hyperparameters of the L-G network. Empirical investigations suggest that setting $\alpha=1$, $\beta=2$, $\gamma=3$, and $\eta=4$ yields optimal results, as detailed in Appendix~\ref{gttvsie}.
Additionally, the Adam optimizer and batch size of 1000 are employed to enhance computational efficiency, training stability, and convergence.
The Huber loss \cite{Huber} is adopted throughout our experiments since it combines the benefits of mean squared error and mean absolute error.
The model's resilience to data and hyper-parameter variations is ensured by implementing multiple experiments and retaining the best-performing model, which can also prevent overfitting and ensure effectiveness in real-world scenarios.

The comprehensive performance evaluation of the trained L-G networks for ${\bar {\cal E}}_{\rm MW}$ and ${\cal V}_{\partial_\theta C}$ is presented in Figs.~\ref{fig2}.
To intuitively showcase their predictive capabilities, confusion scatters are plotted on both training and testing datasets.
Fig.~\ref{fig2}(a) and Fig.~\ref{fig2}(b) respectively illustrate the predictive performance for ${\bar {\cal E}}_{\rm MW}$ and ${\cal V}_{\partial_\theta C}$, where the horizontal axes represent the ground truth values and the vertical axes correspond to the network predictions. The dashed regression lines in both subfigures indicate the ideal prediction scenario, serving as a reference for perfect alignment between predicted and actual values.  
The confusion scatters reveal that the predicted ${\bar {\cal E}}_{\rm MW}$ and ${\cal V}_{\partial_\theta C}$ closely align with their true counterparts yielding root mean square errors (RMSEs) of 0.074 and 0.087 respectively.
The Pearson correlation coefficients (PCs) for both ${\bar {\cal E}}_{\rm MW}$ and ${\cal V}_{\partial_\theta C}$ are also high, as evidenced by their values of 0.95 and 0.93.
Even when handling a wide range of quantum gate counts, the L-G network demonstrates robust learning and generalization performance compared to previously reported neural networks \cite{learnExp1,learnExp2}.
The excellent coincidence of confusion scatters on training and testing datasets suggests the absence of overfitting.

\begin{figure}
\centering
\includegraphics[width=0.48\textwidth]{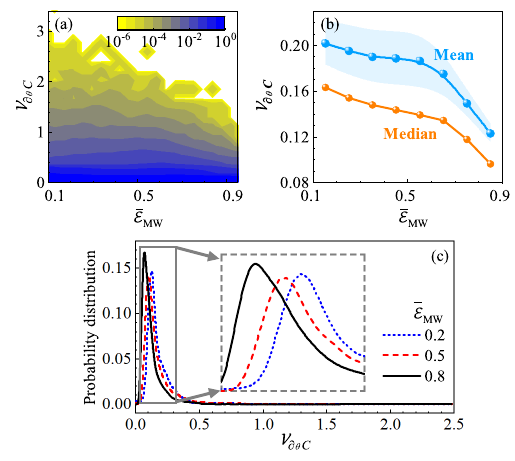}
  \caption{(a) The probability distribution of three million 8-qubit PQCs, each containing 30-50 quantum gates, within the 2D space of ${\cal V}_{\partial_\theta C}$ and ${\bar {\cal E}}_{\rm MW}$. (b) The mean, variance, and median of ${\cal V}_{\partial_\theta C}$ corresponding to each interval of discretized ${\bar {\cal E}}_{\rm MW}$. The blue shadow in Fig.~\ref{fig3}(b) represents the variance of ${\cal V}_{\partial_\theta C}$.
(c) The probability distribution of ${\cal V}_{\partial_\theta C}$ corresponding to ${\bar {\cal E}}_{\rm MW}$ of 0.2, 0.5 and 0.8, respectively.}
  \label{fig3}
\end{figure}

The L-G network for ${\bar {\cal E}}_{\rm MW}$ can be further refined by taking into account the following observations. 
First, the entanglements of PQCs originate from the C-Not gates; 
hence, the equation ${\bar {\cal E}}_{\rm MW}=0$ implies the absence of C-Not gates in PQCs, as evidenced in the leftmost one-column dots in Fig.~\ref{fig2}(a). 
Second, ${\bar {\cal E}}_{\rm MW}$ is a statistical measure and thus should have a non-zero minimum ${\bar {\cal E}}_{\rm MW}|_{\rm min}\simeq0.06$ and a maximum ${\bar {\cal E}}_{\rm MW}|_{\rm max}\simeq0.97$ as long as PQCs incorporate C-Not gates. This is indicated in Fig.~\ref{fig2}(a) and the Appendix~\ref{analyEntData} provides detailed explanations and analysis.
Based on these observations, we exclude PQCs with ${\bar {\cal E}}_{\rm MW}<{\bar {\cal E}}_{\rm MW}|_{\rm min}$ or ${\bar {\cal E}}_{\rm MW}>{\bar {\cal E}}_{\rm MW}|_{\rm max}$ from our analysis.
Note that the aforementioned observations do not apply to ${\cal V}_{\partial_\theta C}$.

\begin{figure}
\centering
\includegraphics[width=0.48\textwidth]{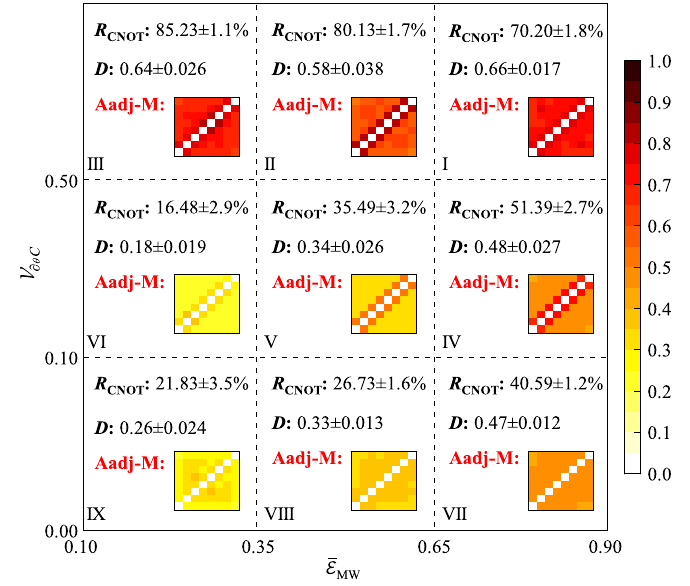}
  \caption{Nine squares divided based on ${\bar {\cal E}}_{\rm MW}$ and ${\cal V}_{\partial_\theta C}$ with 8 qubits and 30-50 quantum gates. In each square, the ratio of C-Not gates ($R_{\rm C\text{-}Not}$) was calculated. PQCs were mapped onto undirected graphs, the circuit connectivity density (D) and the average adjacency matrix (Aadj-M) were calculated for each square.}
  \label{fig4}
\end{figure}

\section{Results} 

\subsection{Theoretical analysis}

To analyze the relationship between ${\bar {\cal E}}_{\rm MW}$ and ${\cal V}_{\partial_\theta C}$, we initially generate three million PQCs and then predict their ${\bar {\cal E}}_{\rm MW}$ and ${\cal V}_{\partial_\theta C}$ using the trained L-G networks.
The main observations are listed below.

\textbf{Observation 1.} \emph{A statistical negative correlation exists between ${\bar {\cal E}}_{\rm MW}$ and ${\cal V}_{\partial_\theta C}$, enabling the circumvention of BP in computational tasks requiring high entanglement.}

Figure \ref{fig3}(a) illustrates the probability distribution of the three million PQCs within the 2D space of ${\bar {\cal E}}_{\rm MW}$ and ${\cal V}_{\partial_\theta C}$.
The contour lines exhibit a downward shift as ${\bar {\cal E}}_{\rm MW}$ increases, except the region where ${\bar {\cal E}}_{\rm MW}\lesssim0.06$, suggesting that trainability exhibits a decreasing trend when PQC's entanglement increases.
This negative correlation between trainability and entanglement is further confirmed by observing the variation in the mean (or median) value of ${\cal V}_{\partial_\theta C}$ with respect to the entanglement, as depicted in Fig.~\ref{fig3}(b).
Such a negative correlation behavior aligns with the earlier reports \cite{18}.
 
It should be pointed out that in Fig.~\ref{fig3}(a) the number of PQCs within each identical interval of ${\bar {\cal E}}_{\rm MW}$ differs.
To counteract this effect, we randomly generate ten thousand PQCs for three cases with ${\bar {\cal E}}_{\rm MW}= 0.2$, 0.5, and 0.8, respectively.
Their probability distributions concerning ${\cal V}_{\partial_\theta C}$ are depicted in Fig.~\ref{fig3}(c).
The peak of the distribution shifts toward the left as ${\bar {\cal E}}_{\rm MW}$ rises from 0.2 to 0.8.
This also underscores the negative correlation between trainability and entanglement, implying an increase in the PQC's entanglement correlating with a higher likelihood of encountering the BP phenomenon.

The aforementioned analysis simultaneously reveals that the negative correlation between trainability and entanglement is a statistical inference.
Thus, it might be possible to construct the PQC with any values of ${\bar {\cal E}}_{\rm MW}$ and ${\cal V}_{\partial_\theta C}$.
To elucidate this, we partition the 2D space of ${\bar {\cal E}}_{\rm MW}$ and ${\cal V}_{\partial_\theta C}$ into nine squares, see Fig.~\ref{fig4}.
By employing the trained L-G networks, we statistically determine three topological parameters of PQCs within each square: the ratio of C-Not gates ($R_{\rm C\text{-}Not}$), the circuit connectivity density ($D$), and the average adjacency matrix (Aadj-M). 
$R_{\rm C\text{-}Not}$ and $D$ are defined as follows,
\begin{align}
R_{\rm C\text{-}Not} = {N_{\rm C\text{-}Not}/L}, \quad
D = {2N_{\rm Edge}}/\left[N(N-1)\right],
\end{align}
where $N_{\rm C\text{-}Not}$ represents the number of C-Not gates and $N_{\rm Edge}$ corresponds to the number of edges in an undirected graph mapped from a PQC \cite{7-34-ent}.
The adjacency matrix is a square matrix representing the connections between nodes in a graph. In this work, Aadj-M is obtained by calculating the average adjacency matrix of PQCs and quantifying the occurrence probability of each qubit pair defined by C-Not gates. Although the precise values of the topological parameters marginally depend on how the nine squares are partitioned, the subsequent findings remain consistent when viewed from a statistical standpoint.

\textbf{Observation 2.} \emph{An appropriate high portion of  $R_{\rm C\text{-}Not}$ can lead to high values of both ${\bar {\cal E}}_{\rm MW}$ and ${\cal V}_{\partial_\theta C}$; however, caution should be exercised to avoid exceeding approximately 70\% which may result in the decrease of ${\bar {\cal E}}_{\rm MW}$.}

It can be seen from Fig.~\ref{fig4} that regarding the low- and moderate-trainability squares (two bottom rows), $R_{\rm C\text{-}Not}$ demonstrates a positive correlation with ${\bar {\cal E}}_{\rm MW}$; however, in the high-trainability squares (top row), $R_{\rm C\text{-}Not}$ exhibits a negative correlation with ${\bar {\cal E}}_{\rm MW}$. The top-right square reveals such threshold value of converting, approximately $70\%$.
Furthermore, $R_{\rm C\text{-}Not}$ attains its smallest values when ${\cal V}_{\partial_\theta C}$ is moderate; whereas $R_{\rm C\text{-}Not}$ reaches its maximum value when ${\cal V}_{\partial_\theta C}$ is high.
Consequently, it can be summarized as Observation 2.

\textbf{Observation 3.} \emph{Increasing $D$ can significantly enhance ${\bar {\cal E}}_{\rm MW}$, but this effect seems to be vague at high ${\cal V}_{\partial_\theta C}$. In circuits with high ${\cal V}_{\partial_\theta C}$, the portion of nearest-neighbor connection is greater than other connections, while in circuits with low ${\cal V}_{\partial_\theta C}$, the portions of various connections tend to be close.}

Concerning the circuit connectivity density $D$, Fig.~\ref{fig4} illustrates a positive correlation between $D$ and ${\bar {\cal E}}_{\rm MW}$. 
However, this correlation can't be verified for high trainability considering the error fluctuation shown by the top row.
Nine Aadj-M heatmaps provide detailed insights into the connections between each pair of qubits. 
The heatmaps in the right column exhibit high connection probabilities, signifying the positive correlation between the complex connection structure and high entanglement. 
A uniformly distributed color represents an equal probability of various connections, as depicted in the bottom row, which naturally correlates with low trainability.
The heatmaps in the top two rows display high-connection probabilities along the secondary diagonal with an offset of 1. 
Consequently, the portion of nearest-neighbor connections of PQCs with moderate and high ${\cal V}_{\partial_\theta C}$ is greater than any other kind of connection.

\subsection{Generating PQCs}

The above findings offer valuable insights for constructing PQCs. Furthermore, leveraging the three topological parameters and the trained L-G networks, we can efficiently filter the randomly generated PQCs to identify the one with specific ${\bar {\cal E}}_{\rm MW}$ and ${\cal V}_{\partial_\theta C}$. 

\begin{algorithm}[H]
  \SetAlgoLined
  \KwData{numbers of gates and qubits, $N$ and ranges of $L$; ranges of ${\bar {\cal E}}_{\rm MW}$ and ${\cal V}_{\partial_\theta C}$.}
  \KwResult{PQC with certain values of ${\bar {\cal E}}_{\rm MW}$ \& ${\cal V}_{\partial_\theta C}$.}
  ranges of $R_{\rm C\text{-}Not}$ and $D$ $\leftarrow$ ranges of ${\bar {\cal E}}_{\rm MW}$ \& ${\cal V}_{\partial_\theta C}$\;
  condition = False\;
  \While{\rm not condition}{
    PQC $\leftarrow$ $L$ and $N$\;
    $R_{\rm C\text{-}Not}$ and $D$ $\leftarrow$ PQC\;
    \If{\rm both $R_{\rm C\text{-}Not}$ and $D$ in their ranges}{
    ${\bar {\cal E}}_{\rm MW}|_{\rm predicted}$ and ${\cal V}_{\partial_\theta C}|_{\rm predicted}$ $\leftarrow$ L-G networks\;
    \If{\rm both ${\bar {\cal E}}_{\rm MW}|_{\rm predicted}$ and ${\cal V}_{\partial_\theta C}|_{\rm predicted}$ in their ranges}{condition = True}}  
    }
  \caption{Generating PQCs.}
  \label{pcode}
\end{algorithm}

The pseudocode is presented in Algorithm~\ref{pcode}. It employs a two-stage screening mechanism:
\begin{enumerate}
    \item \textbf{Initial Screening} (Lines 4-6 in pseudocode): Based on target $\bar{\mathcal{E}}_\text{MW}$ and ${\cal V}_{\partial_\theta C}$ ranges, the algorithm generates a PQC with $L$ gates and $N$ qubits, then evaluates its $R_\text{C-Not}$ and $D$ parameters. Only PQCs with these parameters within specified ranges advance to the next stage.

    \item \textbf{Performance Prediction} (Lines 7-9 in pseudocode): For qualifying PQCs, an L-G network predicts their $\bar{\mathcal{E}}_\text{MW}$ and ${\cal V}_{\partial_\theta C}$ values. A PQC is accepted when its predicted values fall within target ranges.
\end{enumerate}

\begin{figure*}
\centering
\includegraphics[width=1\textwidth]{fig6.pdf}
  \caption{Examples of the PQCs with specific ${\bar {\cal E}}_{\rm MW}$ and ${\cal V}_{\partial_\theta C}$, drawn using \emph{pennylane}. Each PQC, consisting of 30 to 50 quantum gates, is generated by the Algorithm~\ref{pcode} in the main text, with the corresponding search count for PQCs and running time displayed. The predicted and true values of ${\bar {\cal E}}_{\rm MW}$ and ${\cal V}_{\partial_\theta C}$ are presented for each PQC.}
  \label{SI-fig7}
\end{figure*}

This dual-screening approach offers two advantages: first, the topological parameter pre-screening significantly reduces the number of circuits requiring detailed performance prediction; second, the use of trained L-G networks eliminates the time-consuming parameter space sampling process inherent in traditional methods. The algorithm iterates (Lines 3-12 in pseudocode) until finding a circuit that satisfies all criteria.

To illustrate the effectiveness of the Algorithm~\ref{pcode}, we construct a PQC for each square in the nine-square diagram in Fig.~\ref{fig4}.
The generated PQCs are shown in Fig.~\ref{SI-fig7}.
The predicted and actual values of ${\bar {\cal E}}_{\rm MW}$ and ${\cal V}_{\partial_\theta C}$ for each PQC validate the accuracy of the algorithm, while the notably short execution time underscores its high efficiency.

Utilizing the conventional parameter space sampling approach with \emph{Pennylane} and \emph{NumPy} libraries, the computation of ${\bar {\cal E}}_{\rm MW}$ (sampling 20,000 times) for a single PQC requires approximately 200 seconds and the evaluation of ${\cal V}_{\partial_\theta C}$ (sampling 1000 times) takes roughly 400 seconds. However, employing the trained L-G networks reduces this time to just a few or tens of milliseconds.
As shown in Fig.~\ref{SI-fig7}, the number of PQCs searched before finding the target one is in the thousands. 
The conventional sampling approach would necessitate approximately $(200 + 400)\times1000$ s $\approx 7$ days. Nevertheless, employing Algorithm 1 to search for the PQC requires only several to tens of seconds. Consequently, the time efficiency is increased by approximately one million times in searching the required PQC with Algorithm~\ref{pcode}.

\section{Conclusion}

In summary, this work firstly demonstrates the potential of the L-G networks for predicting the entangling capability and trainability of PQCs.
Subsequently, datasets obtained by these networks are employed to explore the relationship between entanglement and trainability, confirming that the negative correlation is statistical and thus can be circumvented with elaborate construction of circuit structures. Further analysis of circuits with both high entanglement and high trainability is conducted with three defined topological parameters of PQCs. It is found that increasing the portion of non-local gates appropriately can increase the entangling capability while its impact on trainability varies with the topological structures of connectivities. The portion of nearest-neighbor connections seems to be positively correlated with trainability.
Finally, using the statistical properties of the topological parameters and the trained L-G networks, we arrive at the algorithm that can quickly and efficiently generate the PQCs with as-required entanglement and trainability.

\section{Acknowledgement}

This work was supported by the National Natural Science Foundation of China (Nos. 61975005 and 12074037), the Beijing Academy of Quantum Information Science (No.Y18G28), and the Fundamental Research Funds for the Central Universities (No.YWF-22-L-938).

\appendix
\setcounter{equation}{0}
\setcounter{figure}{0}
\renewcommand\theequation{A\arabic{equation}}
\renewcommand{\thefigure}{S\arabic{figure}}

\section{The L-G network settings}\label{LGStrcture}

\begin{figure}[h]
\centering
\includegraphics[width=0.38\textwidth]{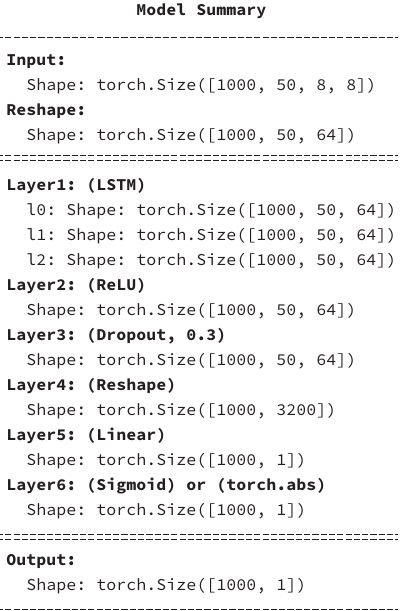}
  \caption{Summary of the structure of the L-G network.}
  \label{SI-fig5}
\end{figure}

Fig.~\ref{SI-fig5} presents the detailed structure of the L-G networks used to predict either ${\bar {\cal E}}_{\rm MW}$ or ${\cal V}_{\partial_\theta C}$. The L-G networks comprise six successive layers designed for sequential data processing and feature extraction. The network accepts input tensors of dimension $(1000 \times 50 \times 8 \times 8)$, which undergo an initial reshaping to $(1000 \times 50 \times 64)$ to facilitate subsequent processing.

The primary feature extraction is performed by an LSTM layer, consisting of three internal states $\{l_0, l_1, l_2\}$, each maintaining dimensions of $(1000 \times 50 \times 64)$. This configuration enables efficient capture of temporal correlations in the input data. The LSTM output is processed through a ReLU activation function, followed by a dropout layer (rate = 0.3) for regularization, both preserving the tensor dimensions. The feature space is then transformed via a reshaping operation to $(1000 \times 3200)$, followed by dimensionality reduction through a linear transformation to $(1000 \times 1)$.

The final layer produces scalar outputs for each sample in the batch. Two separate models are employed for ${\bar {\cal E}}_{\rm MW}$ and ${\cal V}_{\partial_\theta C}$ predictions. The former utilizes a Sigmoid activation function to constrain the output within [0,1], while the latter applies an absolute value function to ensure a positive output, both aligning with the respective physical interpretations of these metrics. Additionally, due to the wide range and highly uneven distribution of ${\cal V}_{\partial_\theta C}$, we apply a logarithmic transformation ${\cal V}_{\partial_\theta C}^{'} = log({\cal V}_{\partial_\theta C} + 1)$ as a preprocessing step. This compresses the high-value region while expanding the spacing in the low-value region. After training, an inverse transformation ${\cal V}_{\partial_\theta C} = exp({\cal V}_{\partial_\theta C}^{'}) - 1$ is applied during prediction to restore the original data.

\begin{figure}[h]
\centering
\includegraphics[width=0.48\textwidth]{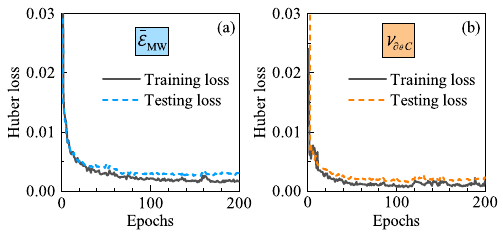}
  \caption{Variations of Huber losses of the L-G networks for predicting ${\bar {\cal E}}_{\rm MW}$ (a) and ${\cal V}_{\partial_\theta C}$ (b) with the training epochs on the training and testing datasets.}
  \label{SI-fig6}
\end{figure}

The learning curves of the L-G networks for predicting ${\bar {\cal E}}_{\rm MW}$ and ${\cal V}_{\partial_\theta C}$ are demonstrated by the Huber losses in Fig.~\ref{SI-fig6}.
Their rapidly downward trends indicate the effectiveness of the L-G networks. 
The minimal deviation in Huber loss between training and test datasets demonstrates the L-G network's robust generalization capability, ensuring the reliability and stability of subsequent analytical results. While further increasing the dataset size would undoubtedly enhance model performance, this choice represents a trade-off between computational efficiency and experimental objectives. As discussed in the main text, generating the dataset is highly time-consuming. Nevertheless, the current dataset size is sufficient to ensure robust results and support the conclusions of this study.

\section{GTT vs IE}\label{gttvsie}

\begin{figure}[h]
\centering
\includegraphics[width=0.48\textwidth]{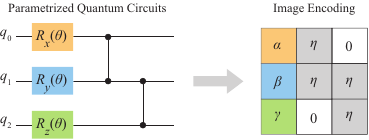}
  \caption{Image encoding (IE) method for parametrized quantum circuits.}
  \label{SI-fig2}
\end{figure}

To encode the circuit structure as input for a machine-learning model, a systematic approach is required to represent circuit structures in the form of tensors. 
The widely used method involves transforming a circuit into a multi-channel image encoding (IE) \cite{2-ent}, as shown in Fig.~\ref{SI-fig2}. 
The tensor has the shape [depth, qubits, gate types], where the depth corresponds to the number of gate layers. 
The size of the image is determined by the number of qubits and the circuit's depth. 
Such an image encoding introduces two constraints on the search space: (a) two-qubit gate operations are exclusively implemented between physically adjacent qubits, and (b) all two-qubit gates should exhibit symmetric action on their respective qubit pairs. These architectural constraints significantly reduce the versatility of the search space.
Compared with the image encoding in Fig.~\ref{SI-fig2}, the GTT encoding (Fig.~1 of the main text) does not require these two conditions.

\begin{figure}[h]
\centering
  \includegraphics[width=0.45\textwidth]{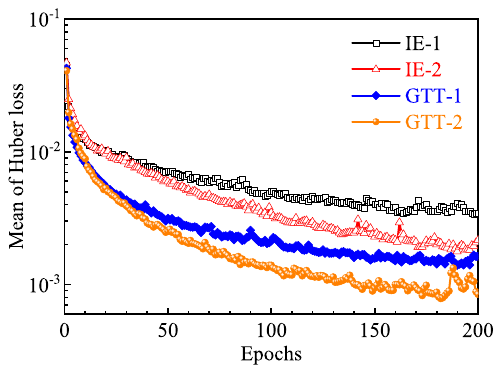}
  \caption{Variation of learning curves with training epochs for predicting ${\bar {\cal E}}_{\rm MW}$ using image encoding (IE) \cite{2-ent} and gate-to-tensor (GTT) encoding. Each curve represents the mean of Huber loss of ten training processes. Encoding parameters: for number 1, $\alpha=10$, $\beta=20$, $\gamma=30$, and $\eta=40$; for number 2, $\alpha=1$, $\beta=2$, $\gamma=3$, and $\eta=4$.}
\label{SI-fig3}
\end{figure}

Figure~\ref{SI-fig3} shows the mean of Huber loss of ten repeated experiments on the testing dataset as a function of the training epoch.
The four cases have identical network structures except for the encoding strategy. 
Both the GTT encodings outperform both the image encodings \cite{2-ent}.
The performance of the GTT encoding depends on the encoding values of $\alpha$, $\beta$, $\gamma$, and $\eta$.
In training, they are taken as the hyperparameters of the L-G network and it is available to take $\alpha=1$, $\beta=2$, $\gamma=3$, and $\eta=4$ after some experiments.

\section{Statistical analysis of ${\bar {\cal E}}_{\rm MW}$}\label{analyEntData}

\begin{figure}[h]
\centering
\includegraphics[width=0.48\textwidth]{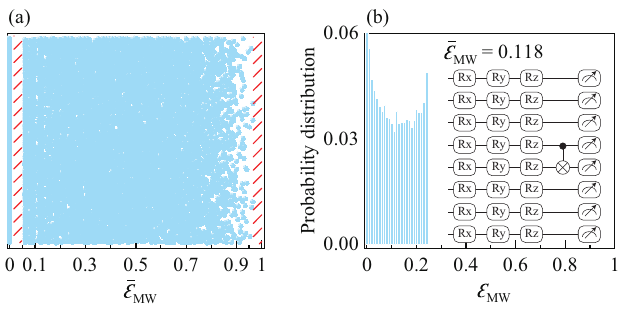}
  \caption{(a) Distribution of PQC's ${\bar {\cal E}}_{\rm MW}$ in the dataset for 8 qubits circuits. There are two gaps near 0 and 1, as indicated by the red diagonal lines. (b) Schematic diagram of calculating ${\bar {\cal E}}_{\rm MW}$ for a PQC with only one C-Not gate using the MW entanglement (${\cal E}_{\rm MW}$) metric. The inset is the corresponding PQC.}
  \label{SI-fig4}
\end{figure}

During experiments, we observed two characteristics for the distribution of ${\bar {\cal E}}_{\rm MW}$. 
Specifically speaking, there are gaps near 0 and 1, as shown in Fig.~\ref{SI-fig4}(a) and Fig.~2(a) of the main text. 
We would like to clarify that the reason for this does not stem from inadequate sampling, whereas it is an inherent aspect of the method used to calculate the entangling capability (${\bar {\cal E}}_{\rm MW}$) of PQCs.
We approximate ${\bar {\cal E}}_{\rm MW}$ by sampling the parameter space and calculating the average ${\cal E}_{\rm MW}$ among the output states of a PQC. 
For the PQC without a C-Not gate, each output state is a product one, resulting in ${\cal E}_{\rm MW}=0$ and consequently, ${\bar {\cal E}}_{\rm MW}=0$. 
As one C-Not gate is added to the PQC, the output states can take the value of ${\cal E}_{\rm MW}$ starting from zero. 
Therefore, averaging ${\cal E}_{\rm MW}$ yields a nonzero ${\bar {\cal E}}_{\rm MW}$, as shown in Fig.\ref{SI-fig4}(b) where ${\bar {\cal E}}_{\rm MW}=0.118$.
Speaking from physics, one C-Not commonly correlates the minimal entanglement, and thus the gap of entanglement near zero is due to averaging, similarly for the gap near 1, referred to Fig.~\ref{SI-fig4}(a).
Note that the minimal non-zero value of ${\bar {\cal E}}_{\rm MW}$ is about 0.06 in Fig.~\ref{SI-fig4}(a), larger than 0.118 in Fig.~\ref{SI-fig4}(b).
This is due to the impact of the C-Not position. 
The maxmum value of ${\bar {\cal E}}_{\rm MW}$ is about 0.97 in Fig.~\ref{SI-fig4}(a).
Here we also point out that these two types of entanglement gaps would, naturally, diminish as the number of qubits increases.

\bibliography{L-G_NN.bib}
\end{document}